\documentclass{iopart}
\usepackage{graphics,graphicx,rotating}
\usepackage[usenames]{color}
\usepackage[percent]{overpic}
\usepackage{mathrsfs,amssymb}

\newcommand{\lie}{\mathcal{L}}

\newcommand{\scri}{{\mycal I}}

\newcommand{\be}{\begin{equation}}
\newcommand{\ee}{\end{equation}}
\newcommand{\beq}{\begin{equation}}
\newcommand{\eeq}{\end{equation}}
\newcommand{\ber}{\begin{eqnarray}}
\newcommand{\eer}{\end{eqnarray}}
\newcommand{\bea}{\begin{eqnarray}}
\newcommand{\eea}{\end{eqnarray}}
\newcommand{\ie}{i.e.}

\DeclareFontFamily{OT1}{rsfs}{}
\DeclareFontShape{OT1}{rsfs}{m}{n}{ <-7> rsfs5 <7-10> rsfs7 <10->rsfs10}{} \DeclareMathAlphabet{\mycal}{OT1}{rsfs}{m}{n}

\begin{document}

{\hfill \footnotesize AEI-2009-043}

\title{Stationary hyperboloidal slicings with evolved gauge conditions}

\author{Frank Ohme$^1$, Mark Hannam$^2$, Sascha Husa$^3$ and \mbox{Niall {\'O}~Murchadha$^2$}}
\address{$^1$ Max-Planck-Institut f\"ur Gravitationsphysik, Albert-Einstein-Institut, Am M\"uhlenberg 1, 14476 Golm, Germany}
\address{$^2$ Physics Department, University College Cork, Cork, Ireland} 
\address{$^3$ Departament de F\'isica, 
  Universitat de les Illes Balears, Cra. Valldemossa Km. 7.5, Palma 
de Mallorca, E-07122 Spain}

\ead{frank.ohme@aei.mpg.de}

\date{\today}

\begin{abstract}
We analyze stationary slicings of the Schwarzschild spacetime defined by members of the
Bona-Mass\'o family of slicing conditions. Our main focus is on the influence of a non-vanishing
offset to the trace of the extrinsic curvature, which forbids the existence of standard 
Cauchy foliations but
at the same time allows gauge choices that are adapted to include null infinity ($\mathscr I$) in
the evolution. These hyperboloidal slicings are especially interesting for observing 
outgoing gravitational waves. We show that the standard 1+log slicing condition admits no 
overall regular hyperboloidal slicing, but by appropriately combining with harmonic slicing, we 
construct a gauge condition that leads to a strongly singularity-avoiding hyperboloidal foliation 
that connects the black hole to $\mathscr I$.
\end{abstract}


\maketitle

\section{Introduction}

The principal motivation driving the development of numerical relativity
has been to treat astrophysically relevant situations in full general
relativity and make predictions for physical measurements --- in particular for future
observations of gravitational waves (GWs). For many years chronic instabilities 
in numerical codes for simulating the coalescence of black-hole binaries posed
a major obstacle for the field, and correspondingly the codes currently used
for black-hole simulations originate from a time when the main focus of the 
field was on the stability of evolution codes, rather than the accuracy of the 
GW content of the numerically generated spacetimes. 
Subsequent developments have lead to the production of waveforms that
are accurate enough for many applications in GW astronomy in the near
future~\cite{Hannam:2009hh}, but succeeding generations of detectors (e.g.\
LISA \cite{LISA1,Danzmann:2003tv} or the Einstein Telescope \cite{etweb}) will however
significantly increase the accuracy requirements for numerical simulations. 

Following the initial breakthrough of 2005 
\cite{Pretorius:2005gq,Campanelli:2005dd,Baker:2005vv},
a wealth of information on the coalescence of black holes has already
been learned from numerical simulations (see
\cite{Pretorius:2007nq,Hannam:2009rd} 
for overviews), and the accuracy of numerical codes has increased
dramatically, see e.g.~\cite{Hannam:2009hh}. If we are to extract the 
maximum possible physical and astrophysical information from GW 
observations, however, we will eventually require yet more accurate, and unambiguous, 
calculations of the GW signals from these systems. 
The calculation of the GW signal is complicated in general relativity 
because observers of signals at astronomical distances from the sources are 
appropriately idealized by quantities defined at null 
infinity~\cite{Penrose:1962ij,Frauendiener:1998yi,Purrer:2004nq}, i.e., 
locations arbitrarily far from the source along lightlike directions. 
In most numerical codes one calculates an approximation
to the GW signal at some set of finite distances (on the order of $\sim100M$, where 
$M$ is the total mass of the system) 
from the source, and extrapolates the result to null infinity the same way
one extrapolates the results from runs at different spacings of the numerical
grid to vanishing grid spacing. This procedure is cumbersome,
computationally expensive, and, often more importantly, 
prone to numerical errors and oddities of the numerical coordinates that
dominate the error estimate of the final signal. 

A preferable approach would be to include null infinity in the numerical 
code, 
and there has indeed been much recent progress in this direction,
for example in characteristic 
evolution~\cite{Babiuc:2005pg,Babiuc:2008qy,Winicour:2008tr}, and the 
hyperboloidal initial-value
problem~\cite{Zenginoglu:2008pw,Zenginoglu:2008uc}. 
In this paper we are concerned with the latter approach, namely constructing 
asymptotically null slices of a black-hole
spacetime, with a view to adapting our methods in the future to the 
construction of hyperboloidal multiple-black-hole initial data for use
in numerical simulations. The construction of hyperboloidal initial 
data is however related to the method that will be used to evolve them,
and in particular to the gauge conditions, which will be the focus of this
article.

The idea of evolving the Einstein equations as a Cauchy problem along
a foliation of spacelike surfaces that reach null infinity
was pioneered by Friedrich \cite{Friedrich83,Friedrich:2002xz}; 
see \cite{Frauendiener:2002iw,Frauendiener04,Husa:2002kk,Husa:2002zc}
for early overviews on numerical work. It was recognized early on that
including null infinity in a numerical simulation naturally
involves a compactification of
an infinite physical domain onto a finite grid, and is prone to lead to singularities
in the differential equations that we wish to solve (for a suggestion
to use more generalized slices see  \cite{Calabrese:2005rs}). Finding a
convenient regularization of 
these equations is one of the major obstacles to the use of null infinity in
numerical simulations.
A point of view that has been advocated
in recent years \cite{Zenginoglu:2008pw,Andersson:2002gn,Husa:2005ns,Moncrief:2008ie} is that
{\em before} formulating the concrete
system of equations to be solved, the geometric structure
of null infinity, which expresses the asymptotically Minkowskian
nature of radiation spacetimes, should be made as manifest as possible in 
the coordinate gauge conditions that will be used.
Such a strategy has worked very well for characteristic evolution 
\cite{Winicour:2008tr,Bishop:1997ik}.
The hope is that this will simplify the
regularization problem, keep the resulting equations as simple as possible, 
and carry over many techniques from
``conventional'' numerical-relativity simulations on finite Cauchy slices.

Before doing this, one must settle on gauge conditions.
It is common practice in numerical relativity, and ubiquitous in current
black-hole-binary codes, to evolve gauge conditions with hyperbolic
equations. For example, the popular moving-puncture method employs 
gauge conditions chosen from the Bona-Mass\'o family \cite{Bona:1994dr}
(see however \cite{Andersson:2001kw,Bonazzola:2003dm,CorderoCarrion:2008nf}
for elliptic-hyperbolic approaches). 

Although one ultimately wants to simulate a dynamical spacetime, stationary
solutions play an important role: in many applications perturbations ultimately
radiate away, and a stationary solution is approached. This is true
in particular for astrophysical black hole spacetimes. It is therefore
fruitful to first think in terms of stationary solutions with respect to the
gauge conditions. In fact, it is somewhat surprising that the search for stationary
representations of black holes which are compatible with the gauge conditions 
one is using was not posed much earlier as one of the fundamental questions in 
numerical relativity.
Stationary solutions also motivate more suitable numerical methods and forms
of initial data. The importance of time-independent descriptions of
Minkowski-spacetime as a starting point for numerical approaches to 
the hyperboloidal initial value problem and as a test-case for the conformal
field equations approach has been discussed in \cite{Husa:2002zc}. 
For recent work
on the associated problem of freezing the coordinate postion of null infinity
in compactified evolutions see \cite{Zenginoglu:2006rj,Zenginoglu:2007jw}.

In \cite{Hannam:2006vv,Hannam:2006xw,Hannam:2008sg} we have discussed explicit
stationary representations describing a 
nonspinning black hole, which are consistent with the specific 
gauge conditions used in the moving-puncture method and represent
a trumpet geometry: the slice
extends from a throat at some finite value $R_0$ of the Schwarzschild radial 
coordinate out to spatial
infinity. In \cite{Zenginoglu:2007jw} such maximally sliced ``trumpet data''
have been matched explicitly to a family of stationary hyperboloidal slices.
In this paper, we construct such ``hyperboloidal trumpet'' slices directly from
an evolved gauge condition, with the hope that it will
provide a useful starting point for developing a method for the
simulation of black-hole-binary spacetimes with asymptotically null
slices. We first study the compatibility of the Bona-Mass\'o family of slicing
conditions \cite{Bona:1994dr} with regular stationary {\em hyperboloidal} slices of 
Schwarzschild-Kruskal spacetime. We find incompatibility for the popular 
``1+log'' subfamily, but compatibility for harmonic slicing and 
certain ``hybrid'' conditions, which we construct to yield stationary slices
which are both hyperboloidal and singularity-avoiding, and which are therefore
interesting regarding the generalization of the moving-puncture method to
the hyperboloidal problem.

\section{Preliminaries}\label{sec:prelim}
\subsection{Geometry of hyperboloidal slices}\label{sec:geometry}

Before embarking on technicalities, let us note that a crucial geometric 
quantity in our work is the (trace of the) 
extrinsic curvature --- in this paper we will choose the convention that the
extrinsic curvature of a 3-dimensional spacelike hypersurface is
defined as the Lie derivative with respect to the timelike unit normal
$n^a$ of the induced
metric $h_{ab}$, 
$K_{ab}=+\frac{1}{2} {\cal L}_{n} h_{ab}$, with a positive sign.
This sign convention is unfortunately not common in numerical relativity, but
we believe it is beneficial for physical intuition: positive mean curvature 
$K=h^{ab} K_{ab}$ then signifies an expanding volume element, while 
negative $K$ signifies contraction.

While we are interested in the asymptotic structure of slices, in order
to simplify notation we will
actually not directly work with the conformal compactification picture 
\cite{Penrose:1962ij}, which has become a standard tool \cite{Wald84},
but we nevertheless 
want to define hyperboloidal slices as regular spacelike hypersurfaces in a 
compactified spacetime in the sense of Penrose
\cite{Penrose:1962ij}. It then follows 
for the physical trace of the extrinsic curvature $K$,
that $K > 0$ at future null infinity $\scri^+$, and 
$K < 0$ at past null infinity $\scri^-$.
The name ``hyperboloidal'' stems from the fact that such
surfaces are analogous to the standard
hyperboloids $t^2 - x^2 - y^2 - z^2 = (3/K)^2$ in Minkowski space,
which provide the simplest example.
A wider class of important standard examples are provided by the
constant mean curvature (CMC) slicings, defined
as $K= \rm const.$, which are important for
cosmology and very well studied in spherical symmetry 
\cite{Malec:2003dq,Malec:2009hg,Gentle2000}. For the purpose of comparing our results to CMC slicing,
it is sufficient to recall that the stationary lapse $\alpha$ for a constant 
$K = \bar K$ in spherical symmetry reads
\beq
 \alpha = \sqrt{ \left( \frac{\bar K R}{3} - \frac{C}{R^2}  \right)^{\hskip-3pt 2} + 1 - \frac{2M}{R} }~, \label{eq:alpha_cmc}
\eeq
where $R$ is the areal radius, $M$ the mass of the Schwarzschild black hole and $C$ an integration 
constant that is of no interest for us.

The key difference between a Cauchy slice and a hyperboloidal slice of an
asymptotically flat spacetime, for the purpose of this paper,
is in the asymptotic behavior of the standard lapse function $\alpha$ and shift
vector $\beta^a$ \cite{York79}. It is well known that for a Cauchy foliation one
has that $\alpha \rightarrow 1$ and $\vert\beta^a\vert \rightarrow 0$ as
$r \rightarrow \infty$, while for hyperboloidal foliations we have that
\begin{equation}
\alpha = \mathcal O(r), \quad \vert \beta^a \vert = \mathcal O(r^2) \quad \mbox{for}~ r \rightarrow \infty,
\end{equation}
where $r$ is a radial coordinate that asymptotically behaves like the 
Schwarzschild radial coordinate.
For a stationary hyperboloidal slice that reaches out to $\scri^+$, 
where the mean extrinsic curvature asymptotes to a positive value, 
the shift vector points inward at large separation, while at  $\scri^-$ 
it points outward. This is intuitively clear from looking at a Penrose diagram
such as figure~\ref{fig:combPenrose}: in order to keep $\scri^+$, say,
at a constant
coordinate location, the shift vector has to point along $\scri^+$, i.e.,
along an ingoing null surface. Likewise, to keep a black hole horizon in
place, an outward pointing shift vector is required, and consequently
the shift vector must be expected to change its sign for stationary slices
that connect a black hole to $\scri^+$ (or a white hole to  $\scri^-$).
Some attention to the direction of the shift vector will be required below.

\subsection{Bona-Mass\'o slicing conditions}\label{sec:bmslicing}

In 1994, Bona \etal~\cite{Bona:1994dr} proposed a family of slicing
conditions that are formulated as evolution equations for the lapse
$\alpha$. Introducing a positive, otherwise arbitrary function
$f(\alpha)$, these conditions can be written as 
\beq
(\partial_t - \lie_\beta ) \alpha = \alpha^2 f(\alpha) (K - K_0)~,  \label{eq:BMslicingK0}
\eeq
where $\lie_\beta$ denotes the Lie derivative along the shift vector $\beta^a$. For $f = 1$ we recover harmonic slicing, $f = n/\alpha$, $n \in \mathbb R$ is called 1+log slicing (with $n=2$ the most popular
choice for black hole evolutions); maximal slicing can be viewed
as corresponding to the limit $f \to \infty$ (see e.g.\ our work \cite{Hannam:2008sg}). 
The general properties of these slicing conditions have been studied in much detail
\cite{Bona:1994dr,Bona:1997hp,Brown:2007tb,Alcubierre:1996su,Alcubierre:2002iq,Cook:1997qc,Hannam:2008sg,Hannam:2006vv}
and they have been implemented in numerical simulations with great
success, in 
particular in the breakthrough simulations that have established the moving puncture
method~\cite{Campanelli:2005dd,Baker:2005vv}. So far these conditions have been used in the
context of asymptotically Euclidean foliations, and the function $K_0$ has usually been
taken as a constant, and explicitly or implicitly been set to zero.

In this paper, we ask the question of whether the slicing condition (\ref{eq:BMslicingK0})
is also useful in the hyperboloidal context. We will still
consider  $K_0$ a constant for simplicity, but we will now take it seriously. 
We find that the value of $K_0$ indeed has a strong influence on the stationary foliations of 
the Schwarzschild spacetime, which for $K_0 \neq 0$ differ considerably from the stationary 
solutions found in
\cite{Cook:1997qc,Hannam:2008sg,Hannam:2006vv,Malec:2003dq}. 

Note first that for $K_0 = 0$ stationary solutions will satisfy
\beq
\lie_\beta \alpha = -\alpha^2 f(\alpha) K~, \label{eq:BMstat}
\eeq
which is indeed consistent with $\alpha\rightarrow 1$,  $\beta^a\rightarrow 0$, 
$K\rightarrow 0$ at infinity, i.e., the condition is consistent with  
asymptotically Euclidean slices.
Looking at the right hand side (RHS) of (\ref{eq:BMslicingK0}), one finds that $K_0 > 0$ locally has a decreasing effect on the lapse,
while $K_0 < 0$ has the effect of increasing the lapse function. The question
arises whether $K_0 \neq 0$ is compatible with stationary slices of the Cauchy type. At first sight
one might naively guess
that  $K_0 > 0$ leads to asymptotically constant-mean-curvature (CMC) slicings that reach $\scri^+$, whereas 
$K_0 < 0$ leads to asymptotically CMC slicings that reach $\scri^-$. We will see that this is indeed the case, when
regular CMC slices exist. 
For an asymptotically Euclidean slice, however, the RHS of (\ref{eq:BMslicingK0}) approaches a constant if $K_0 \neq 0$,
while the left-hand side must go to zero, and therefore
a stationary asymptotically Euclidean slice is incompatible with the choice of $K_0 \neq 0$.
We will now investigate the construction of asymptotically CMC
slices.

\section{Construction of stationary hyperboloidal solutions}\label{sec:construction}

\subsection{1+log Slicing}

We start our detailed analysis with the general version of 1+log slicing and $n=2$, 
\beq
 (\partial_t - \lie_\beta ) \alpha = 2 \alpha (K - K_0)~. \label{eq:1logK0}
\eeq
After introducing our notation, we shall show explicitly that for $K_0 \neq 0$ there are no overall regular time-independent solutions of (\ref{eq:1logK0}) in spherical symmetry that lead to a Cauchy foliation or become asymptotically CMC slices.

The calculations are restricted to the spherically symmetric spacetime of a Schwarzschild black hole with total mass $M$,
\beq
 ds^2 = - \left( 1- \frac{2M}{R} \right) dT^2 + \left( 1- \frac{2M}{R} \right)^{-1} dR^2 + R^2 \: d\Omega^2~,
\eeq
where $T$ and $R$ denote the standard Schwarzschild coordinates and $d\Omega = d\phi^2 + \sin^2 \theta \: d\theta^2$. In the case of stationary (\ie, time-independent) foliations, the trace 
of the extrinsic curvature can be expressed as \cite{Malec:2003dq}
\beq
 K = - \beta' - \frac{2\beta}{R}~, \label{eq:stationaryK}
\eeq
where $'$ denotes the derivative with respect to the areal radius $R$ and 
\beq
\beta = \beta^R/\alpha = \pm \sqrt{\beta^i \beta_i}~. \label{eq:defbeta}
\eeq
Furthermore, the lapse and the shift are related through 
\beq
 \alpha^2 - \beta^2 = 1 - \frac{2M}{R}~. \label{eq:Killingalpha-beta}
\eeq
Thus, time-independent solutions of (\ref{eq:1logK0}) satisfy
\bea
 &\beta \alpha' &= - 2(K - K_0) = 2\beta' + \frac{4\beta}{R} + 2K_0  \label{eq:betaalprim_1log}\\
 \Rightarrow~ &\alpha' &=  \frac{4 R \alpha ^2+6 M+2 R \left(K_0 R
   \sqrt{\alpha ^2+\frac{2 M}{R}-1}-2\right)}{R
   \left(R \alpha ^2-2 R \alpha +2 M-R\right)}~. \label{eq:alprim_1log}
\eea
Note that $\beta$ was defined as the positive root of (\ref{eq:Killingalpha-beta}) to obtain (\ref{eq:alprim_1log}). We could as well have chosen the negative sign, which has the same effect in (\ref{eq:alprim_1log}) as changing the sign of $K_0$. For clarity, we make the following statements for the class of solutions that have positive shift for small $R$. However, it should be kept in mind that swapping the sign both of $\beta$ and $K_0$ has no effect on the solution for the lapse. In
\ref{sec:appendix} we also write down
a coupled system of differential equations for lapse and shift,
which is equivalent to (\ref{eq:Killingalpha-beta}), (\ref{eq:alprim_1log}) 
and where an explicit tracking of the sign of the shift is not necessary.

By closely investigating (\ref{eq:betaalprim_1log}) we immediately convince ourselves that the stationary lapse can neither go to unity, nor can it be CMC-like in the range of large $R$. For $\alpha \to 1$, we find that $\alpha'$, $\beta$ and $\beta'$ vanish in the limit $R \to \infty$, which contradicts (\ref{eq:betaalprim_1log}) for all $K_0 \neq 0$. If $\alpha \simeq R$ and $\alpha' \to \rm const.$, as in the case of the stationary CMC solution (\ref{eq:alpha_cmc}), we can conclude $\beta \alpha' \simeq R$, whereas the RHS of (\ref{eq:betaalprim_1log}) approaches a constant. The only power law that does not lead to a contradiction is 
\beq
\alpha \simeq \sqrt{R} \quad \Rightarrow \quad K \simeq 1/\sqrt{R}
\eeq
and below we will indeed find a family of solutions that show this asymptotic behavior for large values of $R$ and positive $K_0$.

\begin{figure}
 \centering
\begin{overpic}[width=0.5\textwidth]{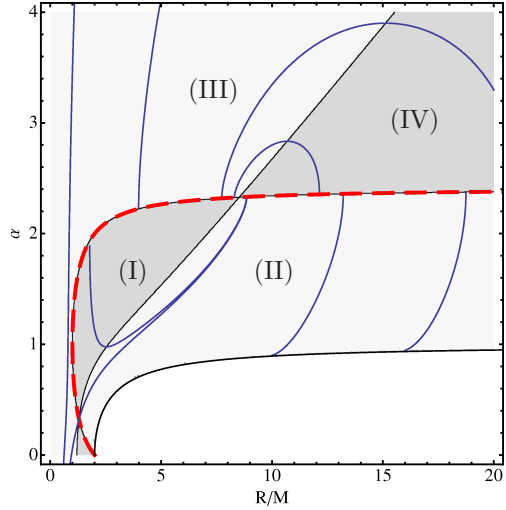}
 \put(21,45){(I)} 
\put(48,45){(II)}
\put(35,81){(III)}
\put(75,75){(IV)}
\end{overpic}
\caption{The analysis of stationary states that satisfy the 1+log equation (\ref{eq:1logK0}) with negative $K_0$ (here, $K_0 = -0.5/M$). $\alpha' > 0$ is indicated by regions (II) and (III), the dark gray regions (I) and (IV) illustrate $\alpha' < 0$. Their boundaries are given by (\ref{eq:denomzero-1log}) (dashed curve), (\ref{eq:numzero-1log}) and $\alpha = \sqrt{1 - 2M/R}$. All solutions become singular where the dashed curve is hit. Some examples which we have calculated explicitly are shown for clarity.}
\label{fig:1log-negK0}
\end{figure}
However, let us first discuss solutions with negative $K_0$ (and positive shift). We analyze the RHS of (\ref{eq:alprim_1log}) by noting that the denominator vanishes along the curve
\beq
 \frac{R}{M} = \frac{2}{1+2\alpha - \alpha^2}~, \label{eq:denomzero-1log}
\eeq
that is characterized by the limit $\alpha \to 1 + \sqrt{2}$ for $R \to \infty$. The numerator of (\ref{eq:alprim_1log}) vanishes for
\beq
 \alpha = \sqrt{1-\frac{3 M}{2 R} + \frac{K_0^2 R^2}{8} - \frac{K_0 \sqrt{R}}{8} \label{eq:numzero-1log}
  \sqrt{K_0^2 R^3+8 M} }~,
\eeq
which grows unboundedly for $R \to \infty$ and $K_0 < 0$.
For every point $(\alpha, R)$, we can easily determine the sign of $\alpha'$ by investigating in which region it is located between the functions (\ref{eq:denomzero-1log}) and (\ref{eq:numzero-1log}).
If a curve that solves the differential equation~(\ref{eq:alprim_1log}) enters a region with negative slope, it is already clear from figure~\ref{fig:1log-negK0} that it will be driven towards the dashed curve (\ref{eq:denomzero-1log}), either directly in region (IV) or by going from (I) to (II). Note that from the discussion above we know that $\alpha'$ cannot vanish in the limit $R \to \infty$. One might think that solutions can ``escape'' to infinity in region (III), but for sufficiently large $\alpha$ and $R$, equation (\ref{eq:alprim_1log}) reveals
\beq
 \alpha' \leq {\rm const.} \left( \frac{4}{R} + \frac{2K_0}{\alpha} \right)~,
\eeq
so that for further increasing $R$ and $\alpha$, all solutions coming from (III) must enter (IV) by crossing the line~(\ref{eq:numzero-1log}).

This finally leads us to the conclusion that every integral curve which satisfies (\ref{eq:alprim_1log}) will inevitably hit the function (\ref{eq:denomzero-1log}) and become singular. Put another way, for negative $K_0$ there exist no smooth solutions of (\ref{eq:alprim_1log}) that are well defined for arbitrarily large $R$.
For completeness, we note that the picture changes for $K_0 < -2.5896/M$, since then the functions (\ref{eq:denomzero-1log}) and (\ref{eq:numzero-1log}) do not intersect at positive $R$ and $\alpha$. However, our arguments are based on the limit $R \to \infty$ and therefore still hold for this case.

\begin{figure}
 \centering
\includegraphics[width=0.5\textwidth]{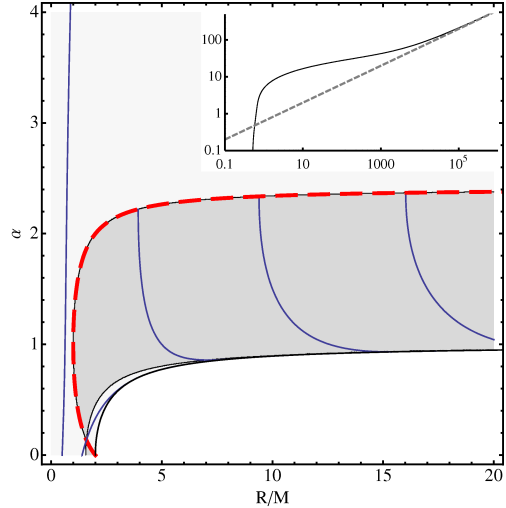}
\caption{The equivalent of figure~\ref{fig:1log-negK0}, but with positive $K_0$ (the chosen example is $K_0 = 0.1/M$). Note that for increasing $R$, the solutions of (\ref{eq:alprim_1log}) are driven away from the dashed curve, to eventually escape to infinity or change the sign of $\beta$ (see text). The included log-log plot illustrates how the outer left curve approaches $\alpha = 2\sqrt{K_0 R}$ (straight dashed line) for large $R$.}
\label{fig:1log-posK0}
\end{figure}
\begin{figure}
 \centering
\includegraphics[width=0.3\textwidth]{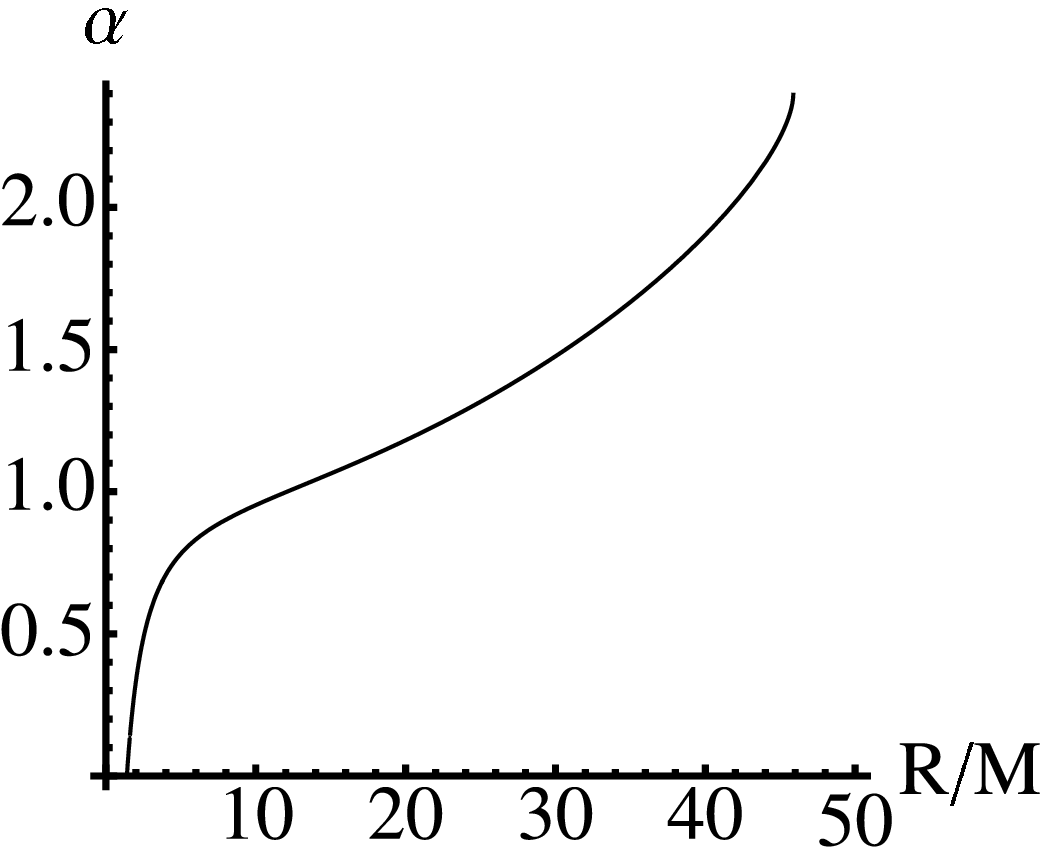} ~
\includegraphics[width=0.3\textwidth]{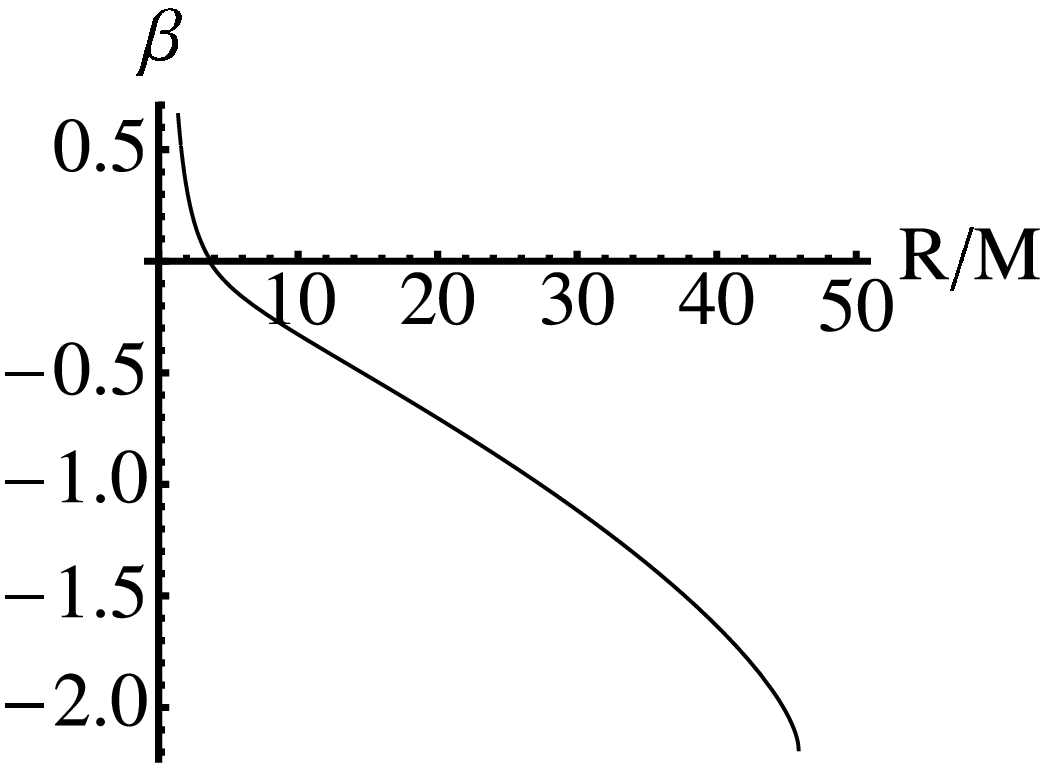} \\[6pt]
\includegraphics[width=0.45\textwidth]{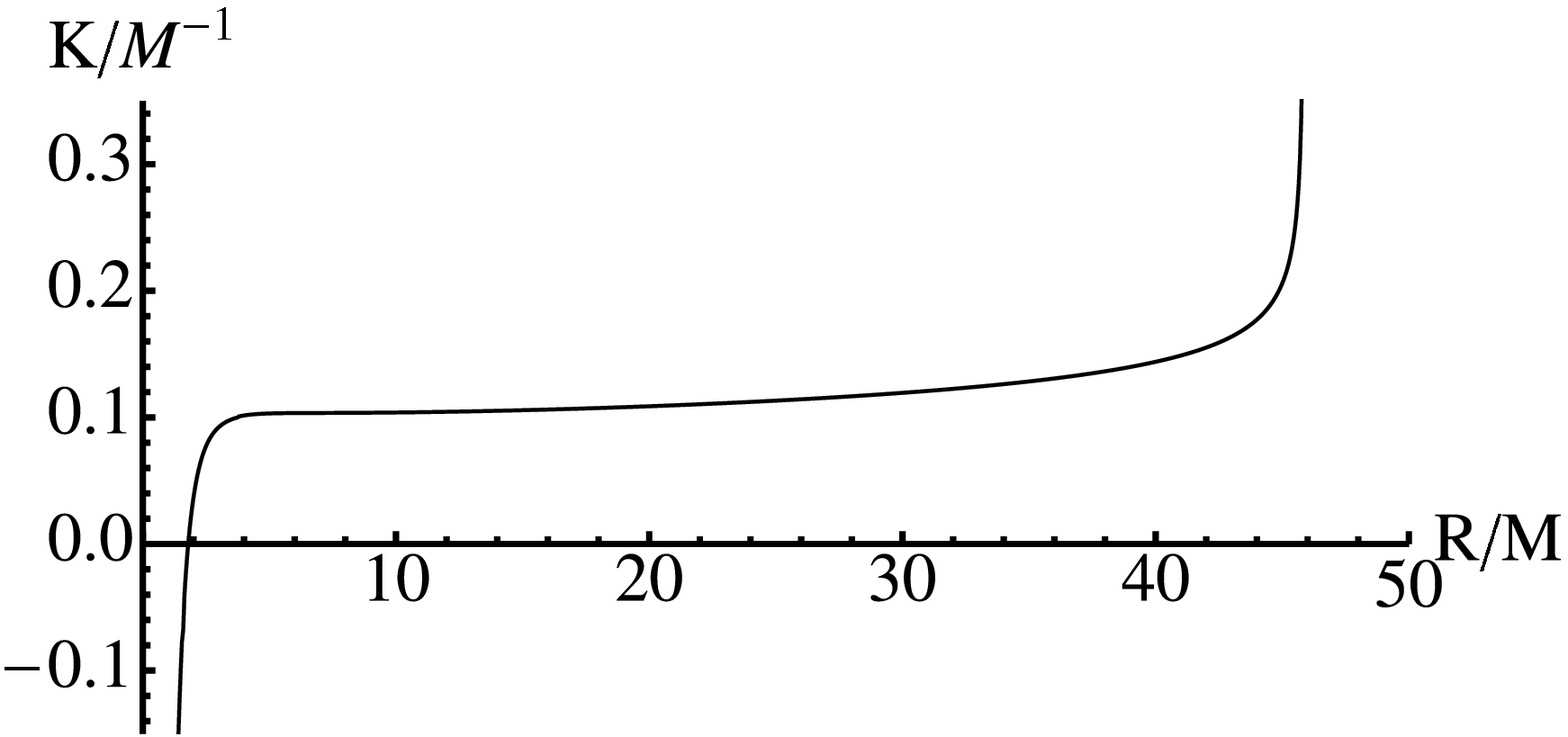} 
\caption{The lapse, the shift and the trace of the extrinsic curvature of the numerically-calculated
critical stationary solution of 1+log slicing with offset $K_0 = 0.1/M$. For small $R$ we recover a behavior similar to the stationary trumpet solution of standard 1+log \cite{Hannam:2008sg}. At $R \approx 3.76M$, however, $\beta$ changes its sign and $K$ seems to settle close to $K_0$, before finally all quantities become singular at $R \approx 46M$.}
\label{fig:1log-criticalSol}
\end{figure}

Let us proceed to positive $K_0$. It is easy to prove that now, the curve of vanishing numerator (\ref{eq:numzero-1log}) approaches the limit $\alpha \to 1$ for $R \to \infty$. Thus, there exist solutions of (\ref{eq:alprim_1log}) that do not become singular but are well defined for arbitrarily large $R$ (and $\alpha > 1+ \sqrt{2}$). We confirmed numerically that these are the predicted solutions with $\alpha \simeq 2 \sqrt{K_0 R}~(R \to \infty)$. Since $\beta$ does not change its sign, the slices are characterized by a negative $K$ and thus connect the black hole to $\mathscr{I}^-$. A visualization of our results is given by figure~\ref{fig:1log-posK0}.
One can also observe that several integral curves which cannot grow unboundedly, hit the boundary of the domain, $\alpha = \sqrt{1 - 2M/R}$. At this point, we find $\beta = 0$, so that (\ref{eq:betaalprim_1log}) is not a well-formulated differential equation any more. However, by exploiting the relation~(\ref{eq:Killingalpha-beta}), we find
\beq
 \beta' = \frac{1}{\beta} \left( \alpha \alpha' - \frac{M}{R^2} \right) \to -K_0
\eeq
for $\alpha \to \sqrt{1 - 2M/R}$. Hence, the shift changes its sign and becomes negative. The curves that hit the boundary of the domain can therefore be smoothly extended by allowing $\beta$ to cross zero at this point. For negative shift, (\ref{eq:alprim_1log}) has to be changed as if $K_0 \mapsto -K_0$.  We therefore conclude that all these solutions run into a singularity at finite $R$. Interestingly, the unique solution that is defined as the one that crosses the common root of (\ref{eq:denomzero-1log}) and (\ref{eq:numzero-1log}) (see for example \cite{Hannam:2008sg,Garfinkle:2007yt,Bruegmann:2009gc} for the discussion of the standard 1+log case) is of this type. Figure~\ref{fig:1log-criticalSol} shows this particular example, which we refer to as the ``critical solution''. As we see, due to the discussed properties of (\ref{eq:alprim_1log}) there is no overall regular critical stationary solution of 1+log slicing with offset $K_0 \neq 0$.

\subsection{Harmonic Slicing}
In this section, we study the harmonic slicing condition
\beq
 (\partial_t - \lie_\beta) \alpha =  \alpha^2 (K - K_0)~. \label{eq:harmonicK0}
\eeq
We find that in this case, $K_0 \neq 0$ does lead asymptotically to CMC solutions at large distances from the black hole. As expected, care has to be taken in assuming the sign of the shift, since for positive $K_0$ it changes from `+' inside the black hole to `$-$' in the limit $R \to \infty$.

The analysis of harmonic slicing can be motivated by considering the generalized Bona-Mass\'o slicing condition (\ref{eq:BMslicingK0}), restricted to $f(\alpha) = \alpha^\kappa$. First, $\kappa$ may be any exponent. Looking for stationary solutions that may asymptotically approach the CMC solution (\ref{eq:alpha_cmc}), we exploit the assumption $\alpha \simeq |\bar K |R/3$ ($R \to \infty$) to determine $\kappa$ and $\bar K$. Starting from (\ref{eq:BMslicingK0}), we find
\bea
 &\beta \alpha' &=  - \alpha^{\kappa + 1} (K - K_0) \\
\Rightarrow ~~ &\frac{|\bar K| \bar K R} 9 &\simeq - \left( K_0 - \bar K \right) \left(  \frac{| \bar K| R} 3 \right)^{\kappa +1} ~.
\eea
By comparing the exponent of $R$ on both sides of the estimate, we conclude that $\kappa = 0$ and 
\beq
 \bar K \to \frac{3K_0}{2} \quad (R \to \infty)~.  
\eeq
Vanishing $\kappa$ obviously leads to the slicing condition (\ref{eq:harmonicK0}).

We now discuss time-independent solutions of (\ref{eq:harmonicK0}) in analogy with our analysis of the 1+log condition
in the previous section. We note that stationary states with positive shift satisfy equivalently to (\ref{eq:harmonicK0}) the following ordinary differential equation,
\beq
 \alpha' =  \frac{\alpha  \left[3 M+R (2 \alpha ^2 - 2 +K_0 R \sqrt{\alpha ^2+\frac{2 M}{R}-1}) \right]}{(2 M-R) R} ~. \label{eq:alprim_harmonic}
\eeq
The set of points of vanishing numerator for $R>0$, $\alpha > 0$ is still described by (\ref{eq:numzero-1log}). The denominator is zero at the horizon, $R = 2M$. In this special case, we can analytically express the regular singular point $(R_s, \alpha_s)$, at which both functions intersect, as
\beq
 \alpha_s = \frac{1}{2} \sqrt{1 + 2 K_0^2-2 K_0\sqrt{K_0^2+1} }~, \quad R_s = 2M~. \label{eq:regular-sing-harm}
\eeq
Note that we face a different situation than in the 1+log case. Here, the curve of vanishing denominator is a vertical line in the $\alpha$-$R$ diagram. Therefore, every solution of (\ref{eq:alprim_harmonic}) that smoothly connects the inside and the outside of the black hole must cross the line of vanishing denominator. To prevent the integral curve to become singular there, the numerator should vanish at the same point, i.e., the unique regular stationary lapse must pass the point given by (\ref{eq:regular-sing-harm}).

\begin{figure}
 \centering
\includegraphics[width=0.3\textwidth]{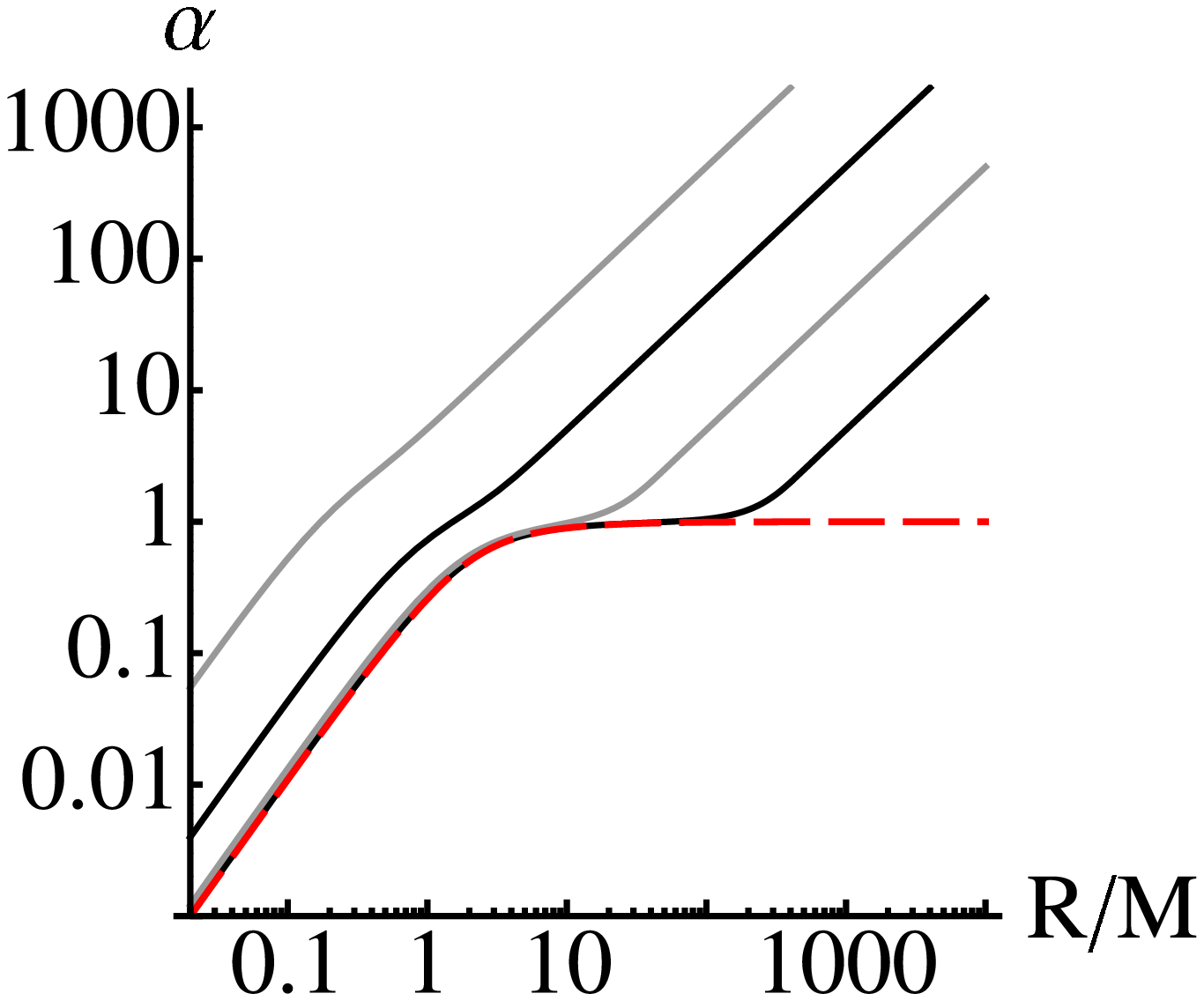}~
\includegraphics[width=0.3\textwidth]{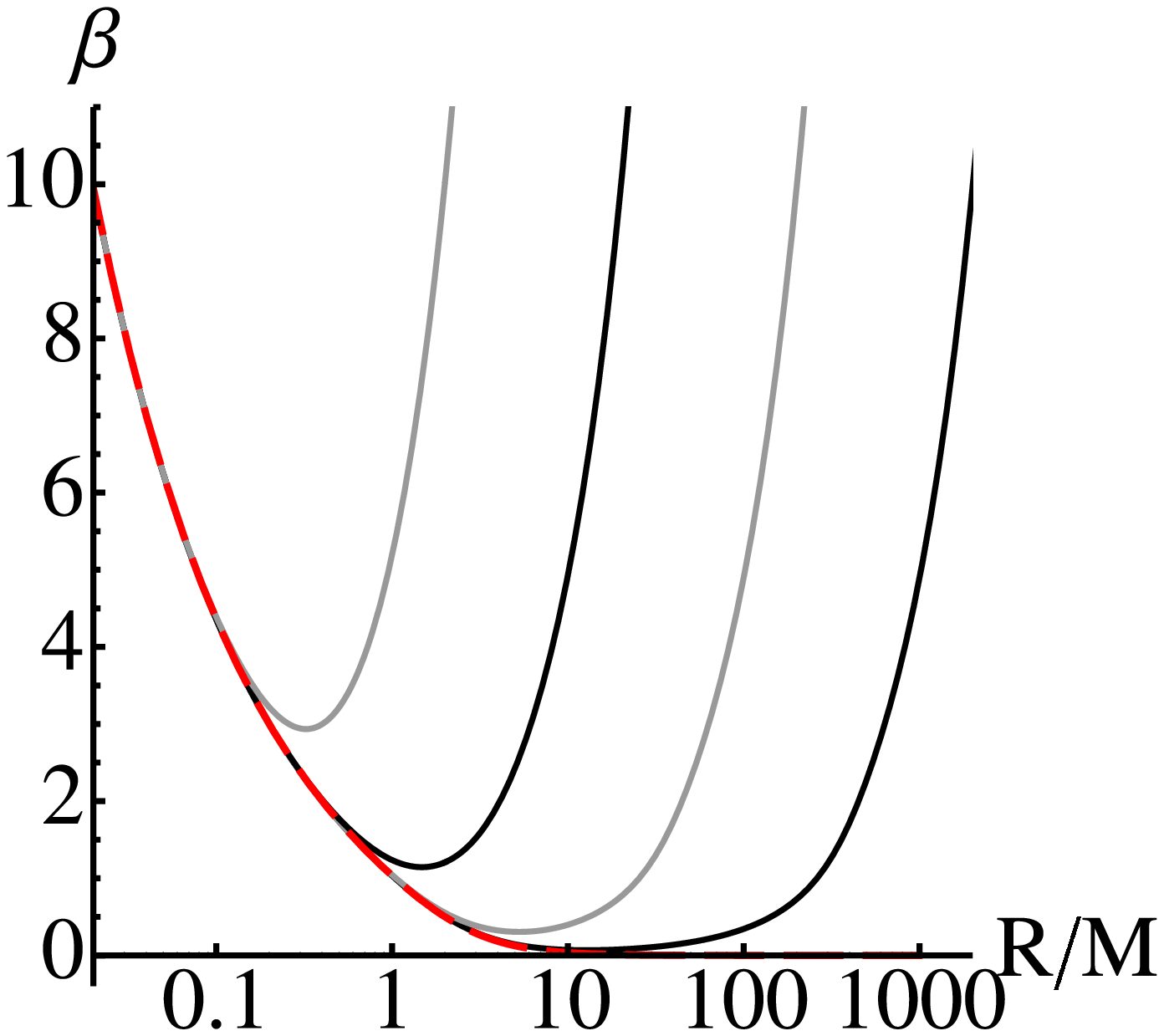}
\caption{The stationary lapse (left) and shift (right) of harmonic slicing (\ref{eq:harmonicK0}). Each curve is obtained by the unique solution of (\ref{eq:alprim_harmonic}) that passes the regular singular point (\ref{eq:regular-sing-harm}). The following cases are displayed (from top to bottom) $M K_0 = -10, -1, -0.1, -0.01, 0$, where vanishing $K_0$ is shown as a dashed line. }
\label{fig:harmLapseNeg}
\end{figure}

\begin{figure}
 \centering
\includegraphics[width=0.55\textwidth]{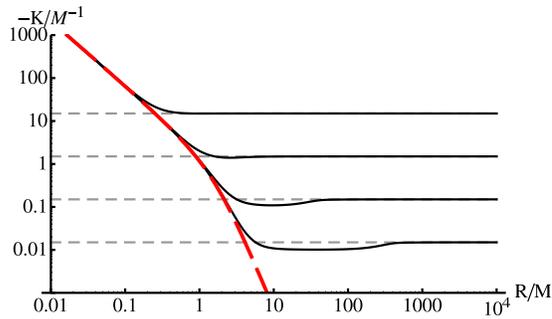}
\caption{The (negative) trace of the extrinsic curvature of the stationary states shown in 
figure~\ref{fig:harmLapseNeg}. The horizontal dashed lines correspond to $3 K_0/2$, the long-dashed curve illustrates the analytically known case of vanishing $K_0$.}
\label{fig:harmKNeg}
\end{figure}

\begin{figure}
 \centering
\includegraphics[width=0.3\textwidth]{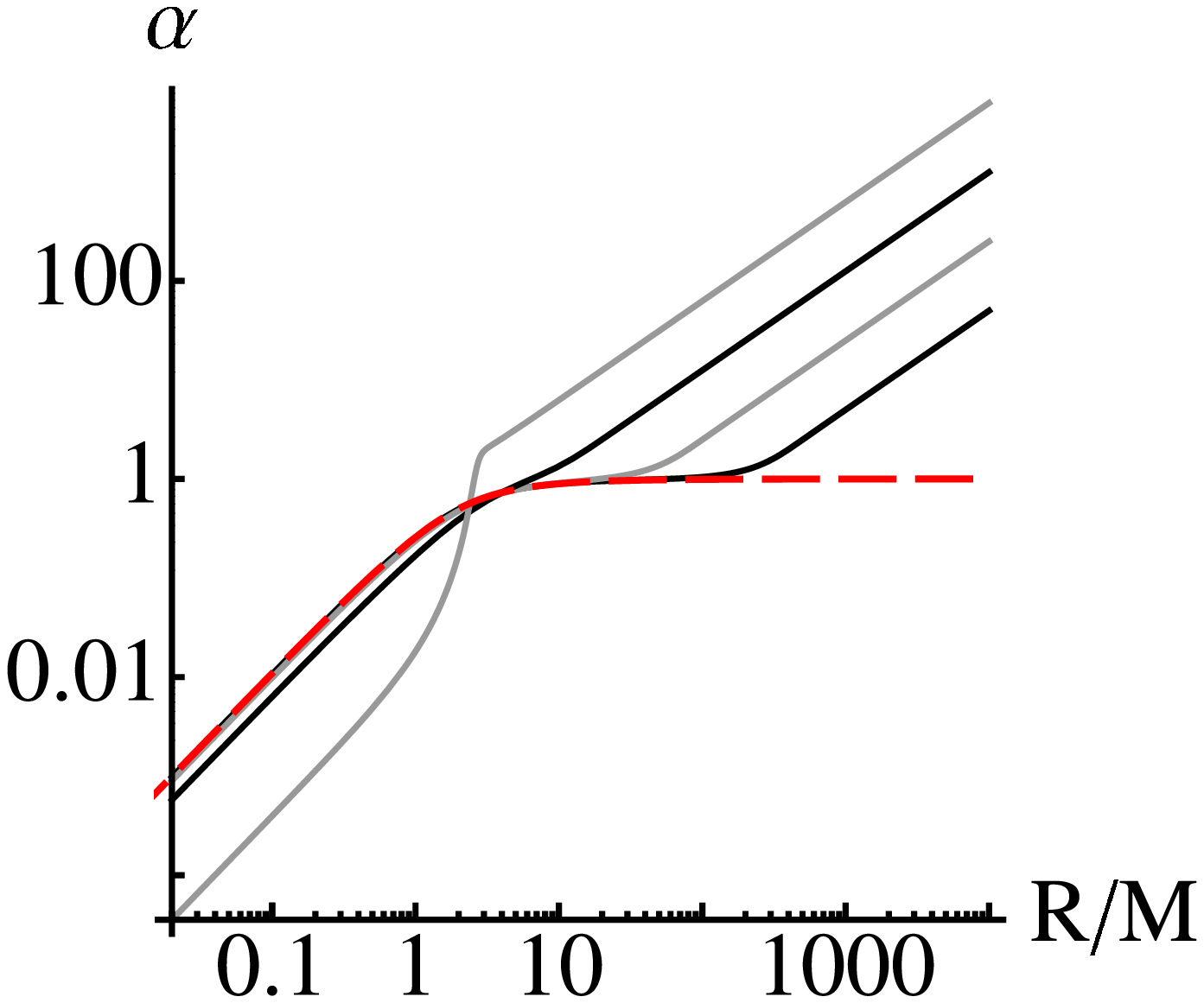} ~
\includegraphics[width=0.3\textwidth]{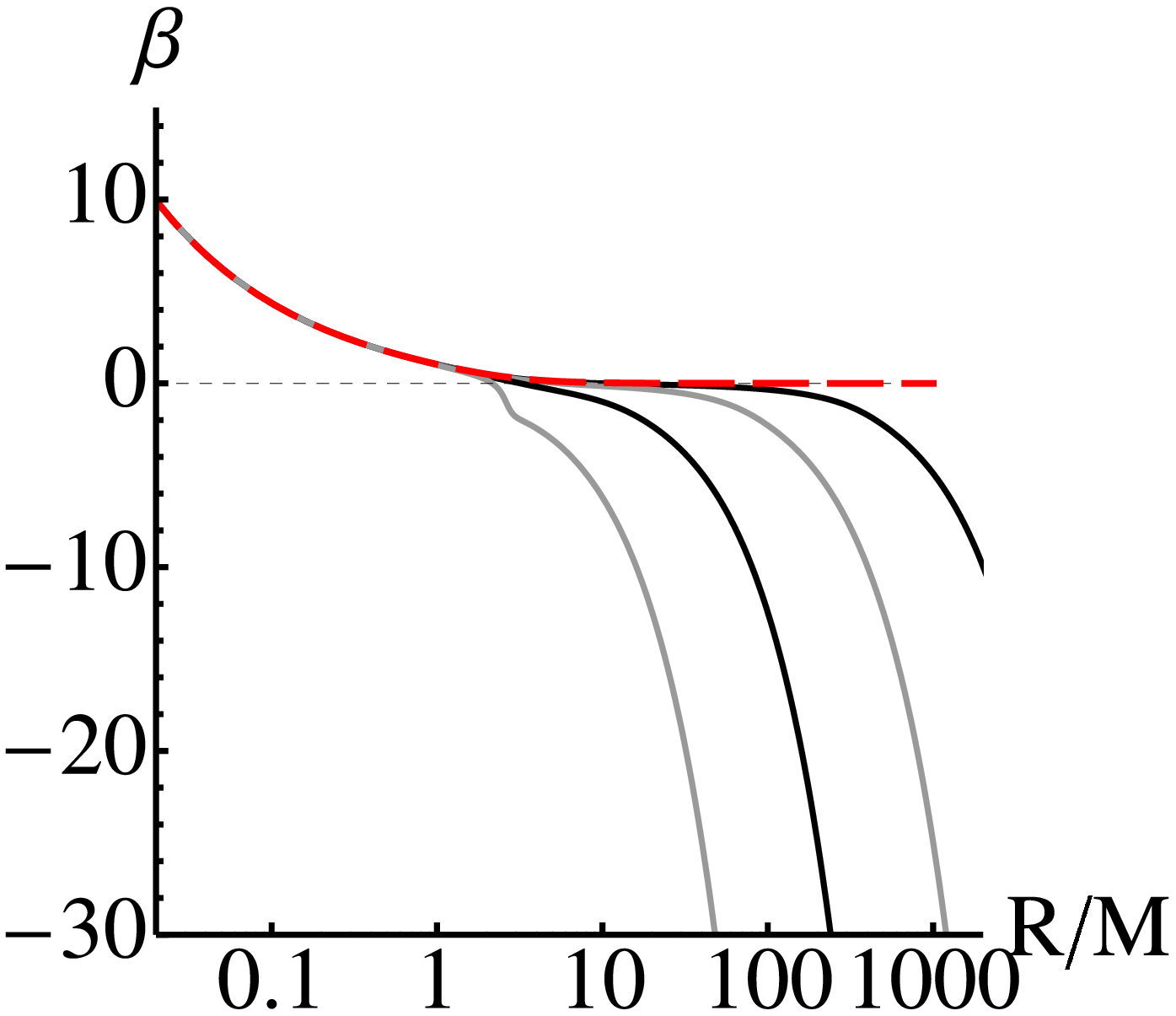} 
\caption{The stationary lapse (left) and shift (right) of harmonic slicing (\ref{eq:harmonicK0}) with non-negative offset $M K_0 = 1.25, 0.25, 0.05, 0.01, 0$ (from top to bottom on the left panel, vice versa on the right panel; $K_0 = 0$ as dashed line).}
\label{fig:harmLapseShiftPos}
\end{figure}

\begin{figure}
 \centering
\includegraphics[width=0.55\textwidth]{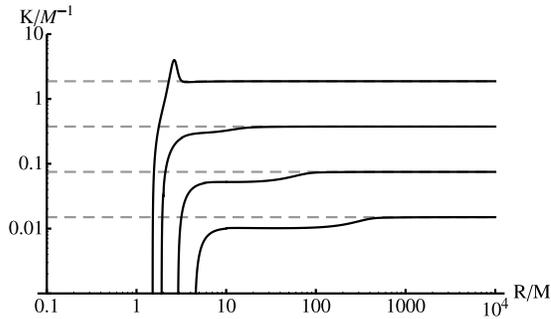}
\caption{The trace of the extrinsic curvature of the stationary solutions that are displayed
in  figure~\ref{fig:harmLapseShiftPos}. The dashed lines correspond to $3 K_0/2$. }
\label{fig:harmKPos}
\end{figure}

We analyze the resulting critical solution for the lapse by numerically integrating (\ref{eq:alprim_harmonic}). Our results show that $K$ indeed asymptotically approaches $3 K_0/2$ for large $R$ in the case of both negative and positive $K_0$. The plots for different choices of $K_0$ are shown in figures~\ref{fig:harmKNeg} and \ref{fig:harmKPos}. Furthermore, the left panels of figures~\ref{fig:harmLapseNeg} and \ref{fig:harmLapseShiftPos} confirm that the lapse asymptotically grows unboundedly similar to CMC slicing, i.e., $\alpha \simeq | K | R/3 \simeq | K_0 | R/2$.

 For negative $K_0$, the shift $\beta$ is decreasing for small $R$ until the $K_0$-dependent minimum is reached, then the slope changes its sign and $\beta \to \infty$ for $R \to \infty$. Thus, the shift is positive everywhere. The trace of the extrinsic curvature remains negative for all $R$. 
We conclude that the stationary foliation of harmonic slicing with \emph{negative offset} connects the singularity $R = 0$ (at infinite proper distance) to \emph{past lightlike infinity} $\mathscr I^-$.%

\emph{Positive offset}, on the other hand, always leads to stationary states where $\alpha$ hits the boundary of the domain, $\alpha = \sqrt{1 - 2M/R}$. $\beta$ vanishes at this points and changes its sign from positive to negative, as described for 1+log in the previous section. By similar calculations one easily proves that again
\beq
 \beta' \Big|_{\alpha = \sqrt{1-\frac{2M}{R}}} = -K_0~.
\eeq
The numerical calculations also suggest that the root of the shift converges monotonically to the horizon for increasing $K_0$, which may lead to problems when numerically solving (\ref{eq:alprim_harmonic}) for large $K_0$. However, as a proof of principle we show several choices $K_0 \leq 1.25 M^{-1}$ in figures~\ref{fig:harmLapseShiftPos} and~\ref{fig:harmKPos}. The important difference to the solutions with negative offset is that here, changing the sign of $\beta$ induces a change of the sign of $K$. Asymptotically approaching a constant \emph{positive} $K$ yields a slicing that smoothly connects the singularity of the black hole with \emph{future null infinity} $\mathscr I^+$.


\subsection{Combined approach}

Let us summarize what we have found in the previous sections. Starting from a general form of Bona-Mass\'o slicing~(\ref{eq:BMslicingK0}), we analyzed the 1+log case ($f = 2/\alpha$). After concluding that in this case there is no overall regular stationary state that connects the black hole to $\mathscr I^+$, we found that harmonic slicing ($f = 1$) is suitable for having asymptotically CMC-like stationary states. The drawback in this case is that harmonic slicing is only marginally singularity-avoiding --- the slicing comes arbitrarily close to the singularity of the black hole. What we propose now is to combine both attractive properties, the strong singularity avoidance of 1+log and the asymptotically constant $K > 0$ of harmonic slicing (with positive offset) by simply adding both conditions in an appropriate way,
\bea
&f = \frac{2}{\alpha} + 1 \\ 
\Rightarrow ~~ &(\partial_t - \lie_\beta)\alpha  = (2 \alpha + \alpha^2)  (K - K_0)~. \label{eq:combinedBMslicing}
\eea
The function $f$ is obviously dominated by the 1+log part for small $\alpha$ (i.e., inside the black hole) and by the harmonic part for large $\alpha$ (which is in the case of CMC slicing equivalent to large distances from the black hole).

We analyze stationary solutions of (\ref{eq:combinedBMslicing}) in the case of a Schwarzschild spacetime by using the established methods and equations $[$recall especially (\ref{eq:stationaryK}), (\ref{eq:defbeta}) and (\ref{eq:Killingalpha-beta})$]$. The time-independent form of (\ref{eq:combinedBMslicing}) can be written as
\beq
 \alpha' =  \frac{(2+ \alpha)  \left[3 M+R (2 \alpha ^2 - 2 +K_0 R \sqrt{\alpha ^2+\frac{2 M}{R}-1}) \right]}{R [2 M-R(1+2\alpha)]} ~. \label{eq:alprim_combined}
\eeq
The critical solution we consider is defined by passing the common root of the numerator (see (\ref{eq:numzero-1log}) that is valid for arbitrary $f$) and denominator
\beq
 \frac{R}{M} = \frac{2}{1 + 2\alpha} \quad \Leftrightarrow \quad \alpha = \frac{2M - R}{2R}~.  
\eeq
Before we further investigate the set of solutions for positive $K_0$, we give some exact analytical expressions for the case $K_0 = 0$.
\begin{figure}
 \centering
\includegraphics[width=0.55\textwidth]{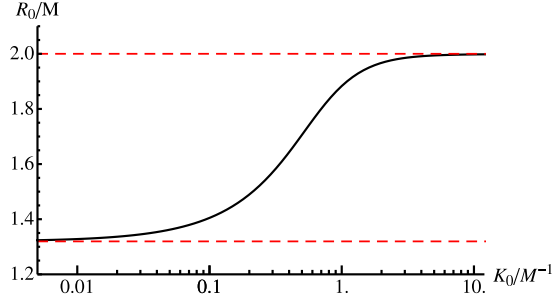}
\caption{The throat (root of the lapse) of the critical stationary solutions of (\ref{eq:combinedBMslicing}). The horizontal dashed lines correspond to the limits $R_0 \to 2M$ for $K_0 \to \infty$ and $R \to 1.3955M$ (see text) for $K_0 \to 0$.}
\label{fig:combR0}
\end{figure}
Under this assumption, time-independent solutions of (\ref{eq:combinedBMslicing}) satisfy
\bea
&\beta \alpha' = (2 + \alpha ) \left(\beta' + \frac{2 \beta}{R} \right)  \nonumber \\
\Leftrightarrow \quad &\int \frac{d\alpha}{2 + \alpha} = \int \left( \frac{\beta'}{\beta} + \frac{2}{R} \right) dR \nonumber \\
\Leftrightarrow \quad &\alpha^2 = 1- \frac{2M}{R} + \frac{C^2 (2 + \alpha)^2}{R^4} ~. \label{eq:combinedImpSol}
\eea
The (positive) integration constant $C^2$ is determined by picking the critical solution as described above. As the regular singular point, we find
\bea
R_s &= \frac{\sqrt{13} +1}{3} M \approx 1.535M~, \\
\alpha_s &= \frac{\sqrt{13} - 3}{4} \approx 0.151~,
\eea
and the constant is given by
\beq
 C^2 = \frac{8}{243} \left(13 \sqrt{13}-35\right) M^4 \approx 0.391M^4~.
\eeq%
The implicit solution (\ref{eq:combinedImpSol}) is now uniquely determined and can also be written explicitly as roots $R(\alpha)$. Note that the roots have to be picked carefully in order to obtain the smooth, monotonically-increasing function we are interested in (see also \cite{Bruegmann:2009gc} for a similar discussion). One important feature is that the resulting lapse starts at $R_0>0 $ with $\alpha(R_0) = 0$ and $\alpha'(R_0) >0$. Hence, the behavior of strongly singularity-avoiding conditions is retained and, expressed in the terminology of \cite{Hannam:2008sg}, the region close to the black hole is a ``trumpet'' with the throat $R_0 \approx 1.3955M$. However, the case $K_0 = 0$ reveals $\alpha \to 1$ for $R \to \infty$, and is therefore not qualitatively different from the standard 1+log.

\begin{figure}
 \centering
\includegraphics[width=0.3\textwidth]{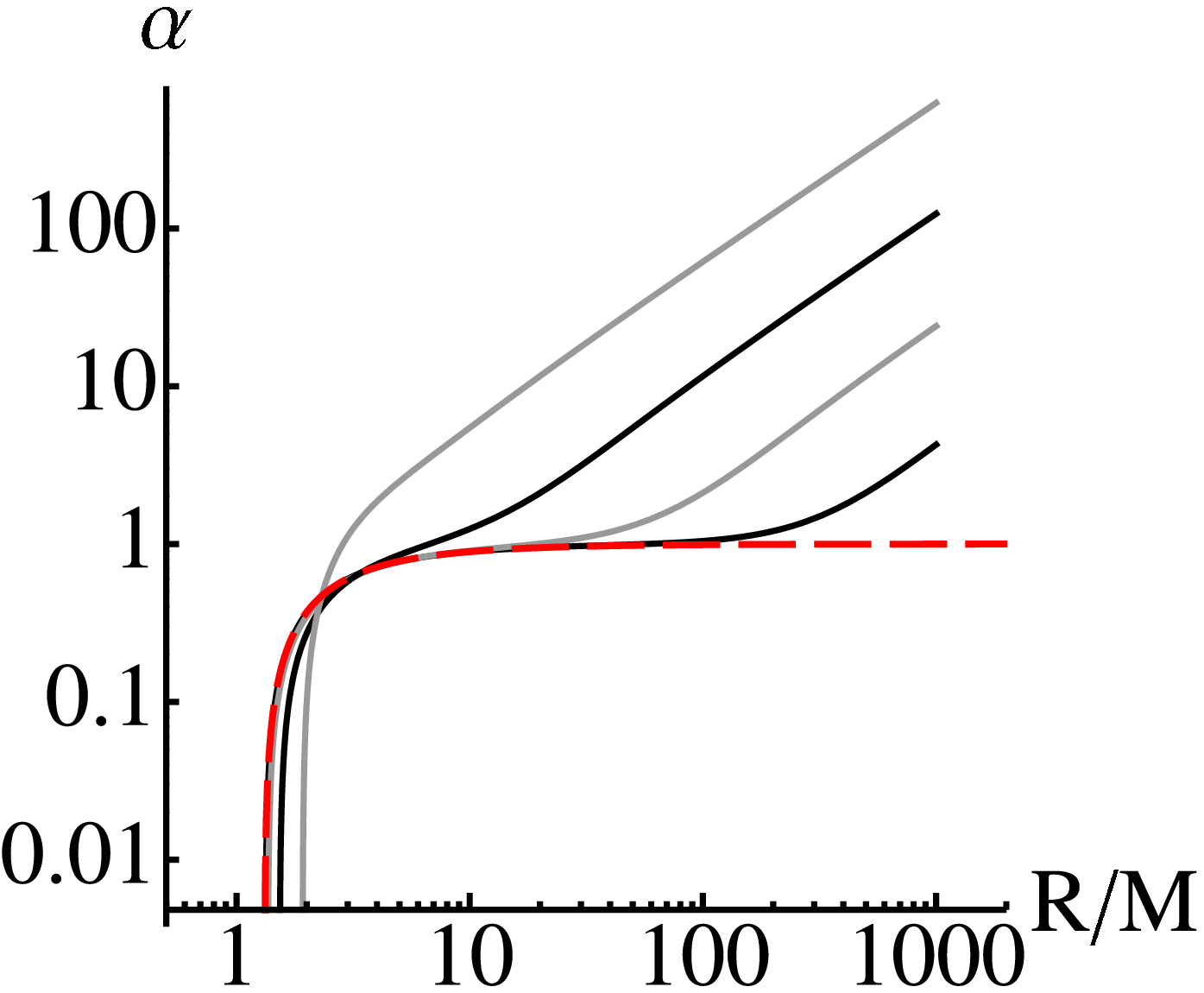} ~
\includegraphics[width=0.3\textwidth]{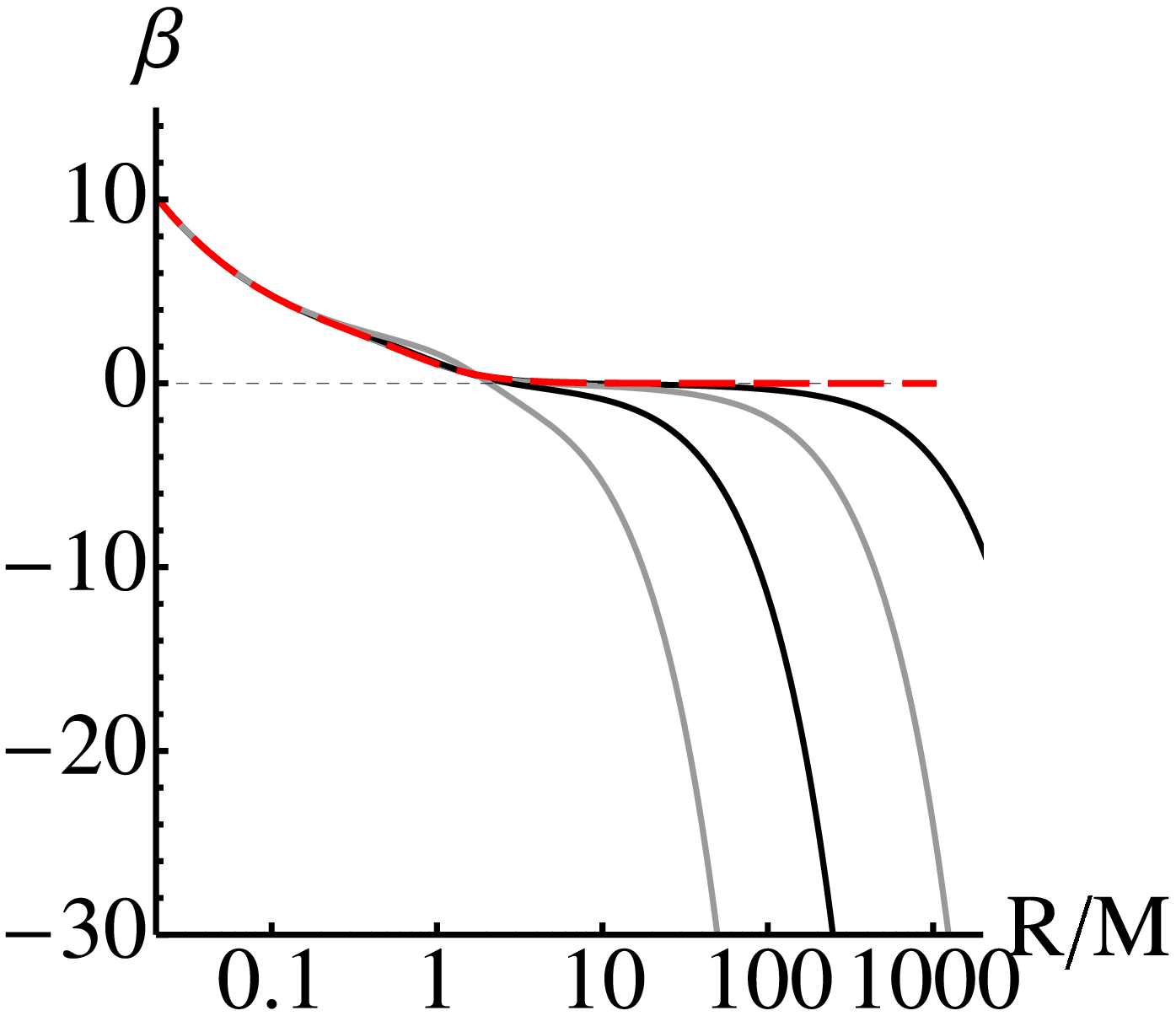} 
\caption{The stationary lapse (left) and shift (right) of our combined slicing condition (\ref{eq:combinedBMslicing}) with non-negative offset $M K_0 = 1.25, 0.25, 0.05, 0.01, 0$ (from top to bottom on left panel, vice versa on the right panel; $K_0 = 0$ as dashed line).}
\label{fig:combLapseShift}
\end{figure}

\begin{figure}
 \centering
\includegraphics[width=0.55\textwidth]{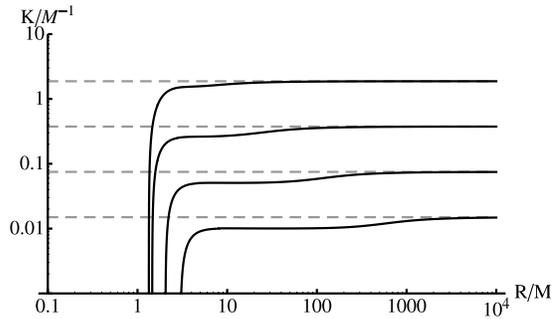}
\caption{The trace of the extrinsic curvature of the stationary solutions that are displayed 
in figure~\ref{fig:combLapseShift}. The dashed lines correspond to $3 K_0/2$. }
\label{fig:combK}
\end{figure}

If we now allow $K_0$ to be positive, we find by numerically integrating (\ref{eq:alprim_combined}) that $\alpha$ still starts at the throat $R_0 (K_0)$ (see figure~\ref{fig:combR0}) but then follows the behavior that was already observed for harmonic slicing, which means that, at finite $R$, the first positive shift changes its sign and the lapse asymptotically approaches the CMC limit $\alpha \simeq K_0 R/2$. The trace of the extrinsic curvature changes sign as well (from negative to 
positive) and approaches the predicted limit $3K_0/2$ for $R \to \infty$. Our results are displayed in figures~\ref{fig:combR0}, \ref{fig:combLapseShift} and \ref{fig:combK}.


\begin{figure}
\centering
\vspace{10pt}
   \begin{overpic}[width=0.6\textwidth]{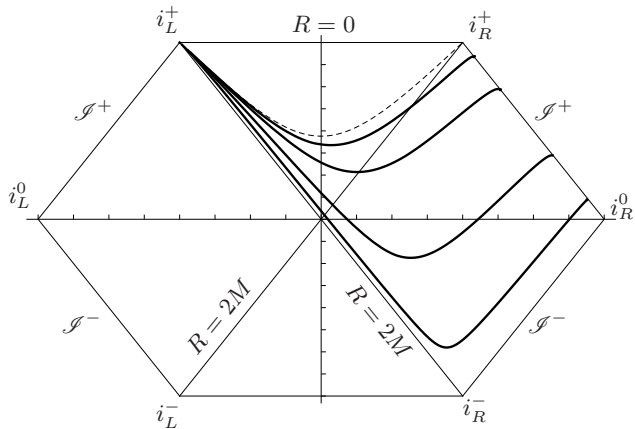}
{\small
\put(75,63){$i_R^+$}
\put(22,64){$i_L^+$}
\put(86,48){$\mathscr{I}^+$}
\put(8,48){$\mathscr{I}^+$}
\put(6,12){$\mathscr{I}^-$}
\put(85,12){$\mathscr{I}^-$}
\put(74,-2){$i^-_R$}
\put(22,-3){$i^-_L$}
\put(-3,34){$i_L^0$}
\put(99,32){$i_R^0$}
\put(29,8){\begin{rotate}{51}
  $R = 2M$
\end{rotate}}
\put(54,19){\begin{rotate}{-51}
  $R = 2M$
\end{rotate}}
\put(45,63){$R = 0$}}
\end{overpic}\vspace{5pt}
\caption{The Carter-Penrose diagram of the slicing defined by the critical stationary solution of (\ref{eq:combinedBMslicing}) and $K_0 = 1M^{-1}$. The dashed line illustrates the curve of constant $R = R_0 \approx 1.8831M$. All slices connect the throat $R_0$ to $\mathscr I^+$. Each displayed time step is $5M$.}
\label{fig:combPenrose}
\end{figure}

In order to point out the new quality of the proposed slicing, we also calculate the embedding in a Carter-Penrose diagram. For this purpose, we transform our integrated quantities $\alpha(R)$ and $\beta(R)$ to standard Schwarzschild coordinates using the height function approach \cite{Malec:2003dq,Beig:1997fp}. Transforming these coordinates to Kruskal coordinates and compactifying them is straightforward and was already described in \cite{Hannam:2008sg}. The resulting diagram is shown in figure~\ref{fig:combPenrose}.
As discussed before, the slices indeed go towards $\mathscr I^+$ (recall, positive $K$) and 
approach timelike infinity $i^+_L$ along the throat, as it is typical for strongly singularity-avoiding slices.

\section{Discussion}\label{sec:discussion}

Let us summarize our results. 

For the 1+log slicing condition used in moving-puncture simulations of black-hole spacetimes,
we have previously found that it is possible to find a stationary solution for a Schwarzschild black hole.
This solution represents a trumpet geometry: the slice
extends from a throat at some finite value $R_0$ of the Schwarzschild radial coordinate out to spatial 
infinity~\cite{Hannam:2006vv,Hannam:2006xw,Hannam:2008sg}. 
In the present paper we have extended these
results to spacelike slices that instead reach null infinity, and which are commonly referred to as hyperboloidal slices.
Such hyperboloidal slices asymptotically approach a finite value of the mean extrinsic curvature, and include
the  constant-mean-curvature slices.
One potential approach to produce trumpet slices that also extend to null infinity  is to modify the 
standard 1+log slicing condition to include an offset term that leads to 
constant-mean-curvature at large distances. We have shown that there are no regular
slices that satisfy this condition, given by (\ref{eq:1logK0}). Also, such an offset
term is an obstruction to construct regular stationary Cauchy slices for the  1+log slicing condition.

If we instead deal with the analogous modification of harmonic slicing, equation~(\ref{eq:harmonicK0}),
regular stationary slices {\it do} exist. However, harmonic slices reach the 
black-hole singularity, and for moving-puncture-like simulations we would prefer
singularity-avoiding slices. Both requirements --- singularity-avoiding 1+log slicing
near the black hole and CMC harmonic slices at null infinity --- can be met by 
using a hybrid slicing condition of the form suggested in (\ref{eq:combinedBMslicing}),
\begin{equation*}
 (\partial_t - \lie_\beta)\alpha  = (2 \alpha + \alpha^2)  (K - K_0)~.
\end{equation*}
We have demonstrated the efficacy of this approach by constructing slices of the 
Schwarzschild spacetime that extend from $R_0$ ($1.3955M < R_0 < 2M$, dependent on the choice of $K_0$) to null infinity.
Both for the hybrid condition and harmonic slicing, the sign of the trace of the extrinsic 
curvature (and therefore the offset $K_0$) is directly related to the outer end ($R \to \infty$) of the slices via
\begin{equation*}
 K \to \frac{3}{2} K_0 \quad \Rightarrow \quad \textrm{slices extend to} \left\{ 
\begin{array}{cl}
 \scri^+ &\quad \textrm{for}~ K_0 > 0 \\
 \scri^- & \quad \textrm{for}~K_0 <0
\end{array}
 \right. ~.
\end{equation*}

There is obviously a large freedom in constructing gauge
conditions and stationary slicings with similar properties, and in choosing
(or generalizing to a function) the constant $K_0$. What works best will
presumably depend to some degree on the application.
 
The results presented in this paper are intended as first steps toward producing
multiple-black-hole ``hyperboloidal trumpet'' data. Numerically evolving such data would 
require a sufficiently regular treatment of null infinity in Einstein's equations. The
philosophy we advocated in the introduction was that the most practical approach 
may be to perform the regularization of Einstein's equations {\it after} choosing the type
of initial data to be evolved, and the gauge conditions to be employed. As such, we
suggest the data and gauge conditions we have presented here as one possible
starting point for deriving a method for null evolutions that is also close in spirit to
current moving-puncture simulations.

%\acknowledgments
\ack
We thank An{\i}l Zengino\u{g}lu, Bernd Br{\"u}gmann and Luc\'ia Santamar\'ia for valuable comments on our manuscript.
Frank Ohme and Mark Hannam thank the University of the Balearic Islands,
and Sascha Husa thanks University College Cork for hospitality,
while some of this work was carried out.
Mark Hannam and Niall \'O~Murchadha were supported by SFI grant
07/RFP/PHYF148.
This work was supported in part by the DFG grant SFB/Transregio~7
``Gravitational Wave Astronomy''. 
S. Husa has been supported in part as a VESF fellow of the
European Gravitational Observatory (EGO), by DAAD grant D/07/13385  
and grant FPA-2007-60220 from the Spanish 
Ministerio de Educaci\'on y Ciencia.

\appendix
\section{Bona-Mass\'o slicing stationary equations as a coupled
system for lapse and shift}\label{sec:appendix}

In the main text we have used a single differential equation for 
the lapse and the quadratic algebraic equation 
(\ref{eq:Killingalpha-beta}) to construct our
slicings --- tracking the solution of 
(\ref{eq:Killingalpha-beta}) with the appropriate 
sign of the shift vector. For completeness, we now also write down
a coupled system of differential equations for lapse and shift, where 
an explicit tracking of the sign of the shift is not necessary.
We write the metric of a Schwarzschild black hole as,
\beq
 ds^2 = - \left(\alpha^2 - \gamma \left(\beta^R \right)^2\right) dt^2 + 
  2 \gamma \beta^R dt \, dR+ \gamma dR^2 + R^2 \: d\Omega^2~,
\eeq
where $t$ is the time coordinate of the stationary slice,  $R$ the standard
Schwarzschild radial coordinate, 
$d\Omega = d\phi^2 + \sin^2 \theta \: d\theta^2$, and $\alpha$, $\beta_R$ and
$\gamma$ are functions of $R$. 
One finds from the Einstein equations that $\gamma = 1/\alpha^2$.
Note that we now use the radial shift component $\beta^R$ instead of $\beta$,
in order to keep track of the direction of the shift.
In the case of time-independent foliations the extrinsic
curvature reads
\beq
 K = -\frac{(r^4 \left(\beta^R\right)^2 \gamma)'}{2 r^4 \alpha \beta^R \gamma}, \label{eq:stationaryK2}
\eeq
where again $'$ denotes the derivative with respect to the areal radius $R$. 
The Einstein equations together with the stationary Bona-Mass\'o slicing
condition (\ref{eq:BMslicingK0})
then imply
\begin{equation}
\alpha' = \frac{f(\alpha ) \left(K_0 r^2 \beta^R +2 r \alpha
   ^3+(3-2 r) \alpha \right)}{r \alpha  (r \alpha  (\alpha -f(\alpha
   ))-r+2)},
\end{equation}
and
\begin{equation*}
\fl {\beta^R}' = \frac{f(\alpha ) \left(2 K_0 r \alpha^3 \left(2 r
   \alpha ^2-r+2\right)+\alpha ^2 \beta^R  \left(7 r \alpha^2-3
   r+4\right)+r \left(\beta^R\right)^3\right)-2 \alpha^3 \beta^R }{2 r \alpha ^3 (r
   \alpha  (\alpha -f(\alpha ))-r+2)},
\end{equation*}
where $r = R/M$.
\section*{References}
\bibliographystyle{unsrt}
\bibliography{hyperb}

\end{document}